\documentclass[fleqn,usenatbib,useAMS]{mnras}
\usepackage{newtxtext,newtxmath}

\usepackage{graphicx}	
\usepackage{amsmath}	
\usepackage{multicol}        
\usepackage{bm}		
\usepackage{pdflscape}	

\usepackage[normalem]{ulem} 
\usepackage[dvipsnames]{xcolor}
\usepackage{soul}

\usepackage{ulem} 
\usepackage{scalerel}

\newcommand{\Ms}{ \mathrm{M}_\odot }

\newcommand{\LCDM}{ $\Lambda$CDM }

\newcommand\Mgii{Mg\protect\scaleto{$II$}{1.2ex}}

\title[The Emperor's New Arc]{The Emperor's New Arc: gigaparsec patterns abound in a $\Lambda$CDM universe}

\author[Sawala et al.]{Till Sawala$^{1,2}$\thanks{Email: till.sawala@helsinki.fi}, Meri Teeriaho$^{1}$, Carlos S. Frenk$^{2}$, John Helly$^{2}$, Adrian Jenkins$^{2}$,  \newauthor 
Gabor Racz$^{1}$, Matthieu Schaller$^{3,4}$, Joop Schaye$^{4}$
\\
$^{1}$Department of Physics, University of Helsinki, Gustaf H\"allstr\"omin katu 2, FI-00014 Helsinki, Finland \\
$^{2}$Institute for Computational Cosmology, Durham University, South Road, Durham DH1 3LE, United Kingdom\\
$^{3}$Lorentz Institute for Theoretical Physics, Leiden University, PO Box 9506, NL-2300 RA Leiden, The Netherlands\\
$^{4}$Leiden Observatory, Leiden University, PO Box 9513, NL-2300 RA Leiden, The Netherlands
}

\date{Accepted XXX. Received YYY; in original form ZZZ}
\pubyear{2022}

\begin{document}
\label{firstpage}
\pagerange{\pageref{firstpage}--\pageref{lastpage}}
\maketitle

\begin{abstract}
Recent discoveries of apparent large-scale features in the structure of the universe extending over many hundreds of megaparsecs, have been claimed to contradict the large-scale isotropy and homogeneity foundational to the standard ($\Lambda$CDM) cosmological model. We explicitly test and refute this conjecture using {\sc FLAMINGO-10K}, a new and very large cosmological simulation of the growth of structure in a $\Lambda$CDM context. Applying the same methods used in the observations, we show that patterns like the ``Giant Arc", supposedly in tension with the standard model, are, in fact, common and expected in a $\Lambda$CDM universe. We also show that their reported significant overdensities are an algorithmic artefact and unlikely to reflect any underlying structure.
\end{abstract}

\begin{keywords}
large-scale structure of Universe, cosmology: theory, methods: numerical, methods: statistical
\end{keywords}




\section{Introduction}
The standard paradigm of hierarchical structure formation predicts that the Universe, while highly inhomogeneous on the scales of galaxies and galaxy clusters, is statistically homogeneous on very large scales \citep{Davis-1985}. Structures extending to several hundred comoving Megaparsecs ($c$Mpc) have long been observed~\citep{Huchra-1982, Gott-1989}. Simulations in the CDM context have since shown that similar features are also predicted to exist in the standard model \citep{White-1987a,Springel-2005-millennium, Park-2012, Schaller-2024} and are, in fact,  required to explain the observed large-scale distribution of galaxies \cite[e.g.][]{Sawala-2024}. It is worth noting that not all previous claims of large-scale inhomogeneities have persisted: the ``Giant Ring of Gamma Ray Bursts" \citep{Balazs-2015} was subsequently shown to be statistically insignificant \citep{Balazs-2018}, while the reported gigaparsec-scale agglomerations of quasars \citep{Clowes-2013} were shown by several authors to be consistent with random fluctuations \citep{Nadathur-2013, Park-2015, Fujii-2024} and compatible with $\Lambda$CDM expectations~\citep{Marinello-2016}.

Recently, the discoveries of two more gigaparsec-scale patterns, the ``Giant Arc" \citep{Lopez-2022} and the ``Big Ring" \citep{Lopez-2024}, have again been interpreted as invalidating the standard model \citep[e.g.][]{Aluri-2023, Lopez-2024b} or as signatures of non-standard physics \citep[e.g.][]{Lapi-2023, Constantin-2023, Mazurenko-2025}. This is partly motivated by the notion of a definite  homogeneity scale of $\approx 370$~Mpc \citep{Yadav-2010} beyond which any structures would be in tension with the standard model.

We use a new, extremely large cosmological simulation of structure formation in $\Lambda$CDM, {\sc FLAMINGO-10K} \citep[Schaller et al.,  in prep.; see also][]{Pizzati-2024}, to test directly the claim that the existence of features like the``Giant Arc'' contradicts $\Lambda$CDM and that its reported overdensity violates the assumption of large-scale homogeneity upon which the standard model is based. 

\section{The``Giant Arc''}\label{sec:GA}
The ``Giant Arc'' was serendipitously discovered from a total sample of 63,876 \Mgii{} absorbers, originally constructed from the catalogue of \cite{Zhu-2013} paired with SDSS DR7 and DR12 quasars, and later identified using a Friends-of-Friends ({\sc FoF}) algorithm \citep{Lopez-2022}. In their analysis, \citeauthor{Lopez-2022} considered a subvolume of $1541 \times 1615 \times 338$ cMpc$^3$ in a redshift interval of $z= 0.742 - 0.862$ containing 504 absorbers. The central redshift, depth of the slice, and linking length of 95 cMpc were all specifically chosen to obtain the clearest identification of the``Giant Arc'' which, with these parameters, contains 44 members.

To ascertain the physical nature and significance of their discovery, \cite{Lopez-2022} applied several statistical tests. They applied the FoF search to 1000 samples with randomised redshifts but did not find another pattern as significant as the ``Giant Arc'', inferring a probability of $< 0.01$ of such a pattern arising at random. Additionally, they computed the convex-hull overdensity of the ``Giant Arc'', and compared it to overdensities of randomised point samples, attributing a $4.5 \sigma$ significance to their structure. We will show that the probability of finding similar structures in both $\Lambda$CDM and in random patterns is much greater than claimed, while the significance of the associated point overdensity is much lower.

\begin{figure*}
\centering
\hspace{-1.5cm}
\includegraphics[width=5.cm, trim={1.4cm 2.3cm 2.9cm 2.0cm},clip]{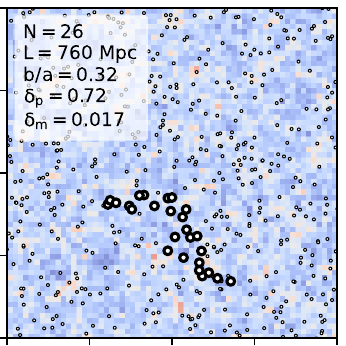} 
\includegraphics[width=5.cm, trim={1.4cm 2.3cm 2.9cm 2.0cm},clip]{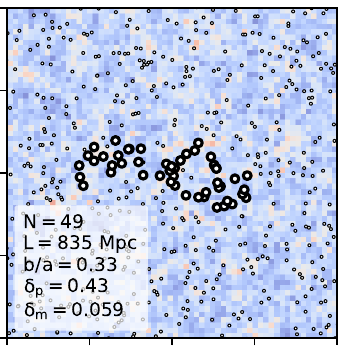}  
\includegraphics[width=5.cm, trim={1.4cm 2.3cm 2.9cm 2.0cm},clip]
{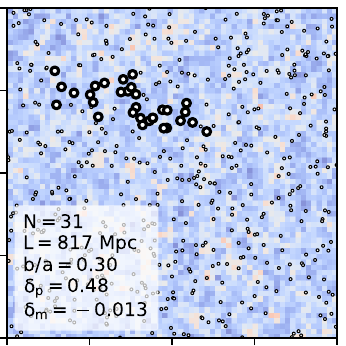}  \\

\hspace{.2cm}\includegraphics[width=5.cm, trim={1.4cm 2.3cm 2.9cm 2.0cm},clip]{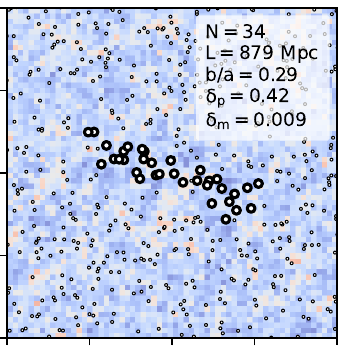} 
\includegraphics[width=5.cm, trim={1.4cm 2.3cm 2.9cm 2.0cm},clip]{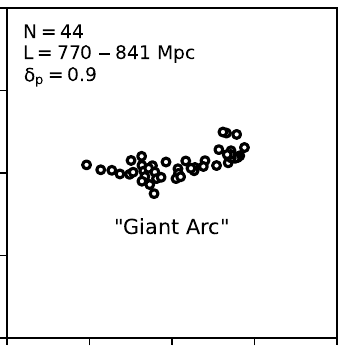}  
\includegraphics[width=6.6cm,  trim={1.4cm 2.3cm 1.0cm 2.0cm},clip]{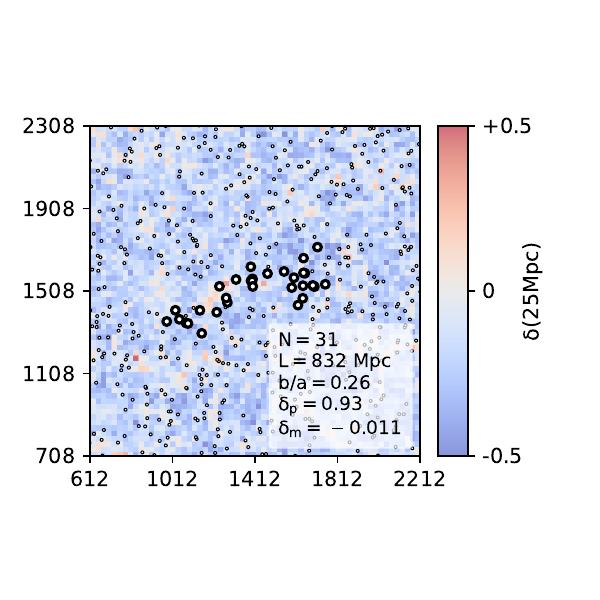} \\

\hspace{-1.5cm}
\includegraphics[width=5.cm, trim={1.4cm 2.3cm 2.9cm 2.0cm},clip]{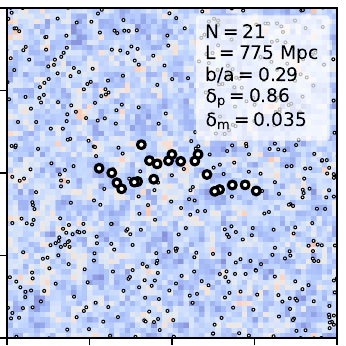} 
\includegraphics[width=5.cm, trim={1.4cm 2.3cm 2.9cm 2.0cm},clip]{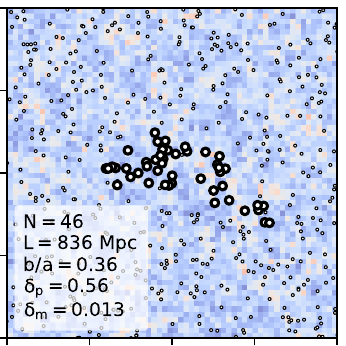} 
\includegraphics[width=5.cm, trim={1.4cm 2.3cm 2.9cm 2.0cm},clip]{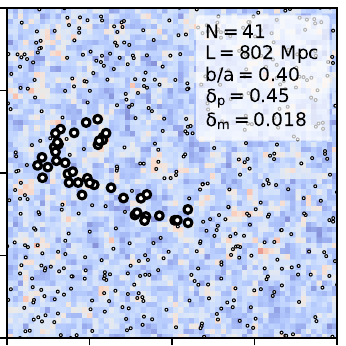}

\caption{A selection of prominent structures identified in slices of the {\sc FLAMINGO-10K} simulation in a single sampling at $z=0.7$, using the same depth, points density, and linking length as in \protect\cite{Lopez-2022}, alongside the``Giant Arc'' reproduced from \protect\cite{Lopez-2022}. Each panel shows a region of $1600 \times 1600$~$c$Mpc$^2$ with a depth of 338~$c$Mpc. The background of each simulation panel shows the projected matter surface overdensity within the slice on scales of 25~$c$Mpc. Subhaloes are overlaid as white circles, with members belonging to the largest structures highlighted. For each structure, the number of points, $N$, maximum extent, $L$, anisotropy, $b/a$, and the convex-hull overdensity of points, $\delta_p$, and of matter, $\delta_m$, are quoted. Nearly every sampling of {\sc FLAMINGO-10K} contains many structures that resemble or exceed the ``Giant Arc'' in size and overdensity. \label{fig:gallery}}
\end{figure*}

\section{The {\sc FLAMINGO-10K} simulation}\label{sec:simulations}
To identify possible counterparts to the``Giant Arc'' in $\Lambda$CDM, we consider {\sc FLAMINGO-10K}, a simulation that follows $10080^3$ collisionless CDM and $5600^3$ neutrino particles ($1.2\times10^{12}$ particles in total) in a periodic volume of $2.8^3$~cGpc$^3$, with cosmological parameters $h=0.681$, $\Omega_\mathrm{m} = 0.306$, $\Omega_\mathrm{\Lambda} = 0.694$, $\Omega_\mathrm{b} = 0.0486$, $n_s = 0.967$, and $\sigma_8 = 0.807$. {\sc FLAMINGO-10K} was performed using the {\sc SWIFT}~simulation code \citep{SWIFT} and is part of the FLAMINGO suite (see \cite{Schaye-2023} for a complete description), using the same phases as their flagship L2p8\_m9 hydrodynamical simulation but with 8$\times$ more collisionless particles.

For comparison to the``Giant Arc'', we use the output at $z=0.70$, the closest for which the full particle data are available, but we also examine different outputs between $z=3$ and $z=0$. From the outputs, self-bound subhaloes were identified in the simulation using the {\sc HBT+} (Hierarchical Bound-Tracing) algorithm (\citealt{Han-2018}; Forouhar Moreno et al., in prep.). As analogues of the \Mgii{} absorbers, we consider subhaloes in the mass range $\mathrm{M_{200,c}} = 1$--$5 \times 10^{12}\Ms$, of which {\sc FLAMINGO-10K} contains $\approx 2.69 \times 10^7$ at $z=0.7$. This mass range brackets results of studies based on clustering strength at similar redshifts \citep[e.g.][]{Lundgren-2011, Gauthier-2014}. The precise halo masses of \Mgii{} absorbers are not known, but as we will discuss, our results are not sensitive to the assumed mass.

The number density of subhaloes in this mass range in {\sc FLAMINGO-10K} is $\approx 1.23 \times 10^{-3}$~$c$Mpc$^{-3}$, about $2000 \times$ the density of \Mgii{} absorbers in the slice of \cite{Lopez-2022}, which were selected on the basis of a (random) association with background quasars. To mimic this selection, we draw random subsamples of subhaloes with the same number density. To provide a baseline for the amount of structures found in the simulation, we follow \cite{Nadathur-2013} and construct random (Poisson) point patterns of equal density.

\section{Structure Finding}\label{sec:methods}
To identify extended structures in the simulation, we follow \citeauthor{Lopez-2022} and apply a Friends-of-Friends algorithm \citep{Huchra-1982, Davis-1985}, i.e. a hierarchical single-link clustering algorithm with a Euclidean distance metric and the linking length as a free parameter. In the more familiar application of finding collapsed objects in cosmological simulations, a canonical linking length of $0.2 \times$ the mean interparticle distance is used. In general, the amount and type of structures found using a {\sc FoF} algorithm are very sensitive to the relation between the linking length and the interparticle distance: a linking length that is too small results in the breakup of structures, while a linking length that is too large results in points being linked together independently of any underlying correlation.

The redshift slice examined in \cite{Lopez-2022} has a volume of $V \approx 0.841$~Gpc$^3$ and a mean point density of $N/V \approx 599$~Gpc$^{-3}$, corresponding to a mean interparticle distance $\left< r \right> = (V/N)^{1/3} \approx 119$~$c$Mpc. The fact that both the mean interparticle distance and the linking length (95 $c$Mpc) are similar to the shortest dimension of the slice (338 $c$Mpc), but much smaller than its other two dimensions makes the geometry of the slice an important factor. Most points are separated by less than one linking length from either face of the slice, and any structures resembling the ``Giant Arc'' will be much longer than its shortest dimension. This largely restricts such structures to two dimensions, which also makes them, by definition, highly anisotropic.

For a direct comparison, we must therefore not only adapt the same redshift, point density and linking length, but also consider equally thin slices of the {\sc FLAMINGO-10K} volume (or random point patterns). Specifically, we examine slices containing 1588 points within $2800 \times 2800 \times 338$~$c$Mpc$^3$ along each of the three orientations of the {\sc FLAMINGO-10K} volume. We exclude structures separated by less than one linking length from any side of the box.

To characterise identified structures, we consider the membership number, $N$, and compute their {\it extent or length, $L$}, which we define as the maximum pairwise point distance. We also compute the {\it point overdensity, $\delta_p$}, for which we calculate the density of the convex hull that encloses spheres of radius equal to the mean distance of the points belonging to the structure, centred on all member points \citep{Clowes-2012}. In addition, we use the simulation particle data to compute the {\it matter overdensity, $\delta_m$}, within the same convex hulls.

We characterise the shape or {\it anisotropy} of structures according to the ratio between the square roots of the second-largest and largest eigenvalues of their covariance matrix, $b/a$. Structures that extend along only one dimension, i.e. arcs or filaments, have one dominant eigenvalue, i.e. $c/a < b/a \ll 1$ (due to the particular geometry, for the extended structures we discuss here, $c \ll a$ by construction).

For the``Giant Arc'', only the membership number $(N=44)$ and overdensity ($\delta = 0.9$) are reported, the latter computed in the same way as our point overdensity, $\delta_p$. The extent of the``Giant Arc'' is only described as ``approximately 1~Gpc''. Based on \cite{Lopez-2022}, we are able to measure its projected extent as $770$~$c$Mpc. Because its depth is at most equal to that of the slice, 338~$c$Mpc, its true extent, by our definition, is at least $770$~$c$Mpc and at most $841$~$c$Mpc.

\section{Structures in {\sc FLAMINGO-10K}}
We find that``Giant Arc'' -like structures are very common features in {\sc FLAMINGO-10K}. In Figure~\ref{fig:gallery}, we show the``Giant Arc'' alongside eight examples of similarly extended structures found in the first random sampling $($seed $= 0)$ of the {\sc FLAMINGO-10K} simulation. All eight structures are disjoint from one another, i.e. no point belonging to any structure is contained in any other. All eight have extents greater than 750~$c$Mpc, contain more than 20 members, and all have point overdensities, $\delta_p$, above 0.4. While no structure in {\sc FLAMINGO-10K} is an exact copy of the``Giant Arc'' (just as no two structures are exact copies of one another), the``Giant Arc'' evidently has many counterparts in a {\sc FLAMINGO-10K}-like $\Lambda$CDM volume.

\begin{figure}
\centering
\includegraphics[width=8.5cm, trim={0.3cm 0.2cm 0.2cm 0.2cm},clip]{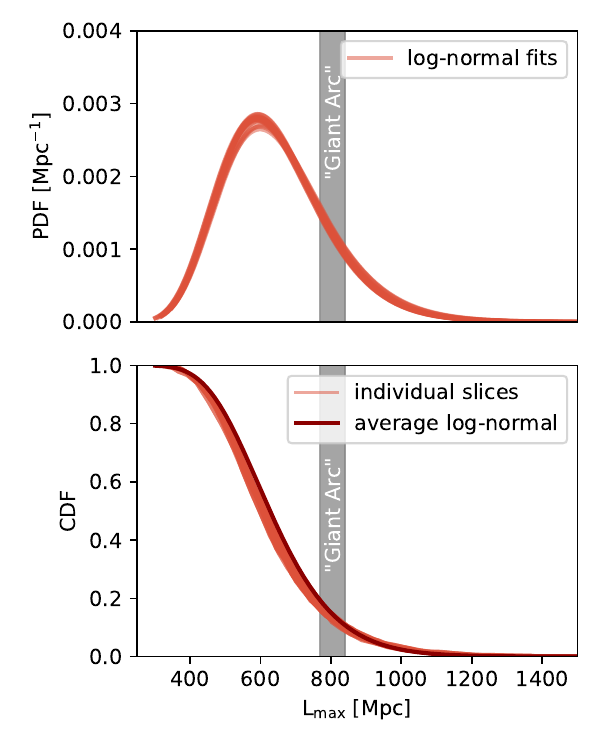}
\caption{Probability density (PDF, top) and cumulative distribution (CDF, bottom) of the lengths of the longest structures with $b/a < 0.5$ in random subsamples within 21 non-overlapping slices. The top shows log-normal fits to the individual PDFs for the 21 individual slices. The bottom panel shows the raw CDFs for each slice, and a log-normal fit to the average PDF. The grey band indicates the lower and upper limit of the extent of the ``Giant Arc'' ($770 \text{$c$Mpc} <L_{\text{max}}<841$$c$Mpc). 
\label{fig:subsamples}}
\end{figure}

\begin{figure}
\centering
\includegraphics[width=8.5cm, trim={1.2cm 1.5cm 0.4cm 1.2cm},clip]{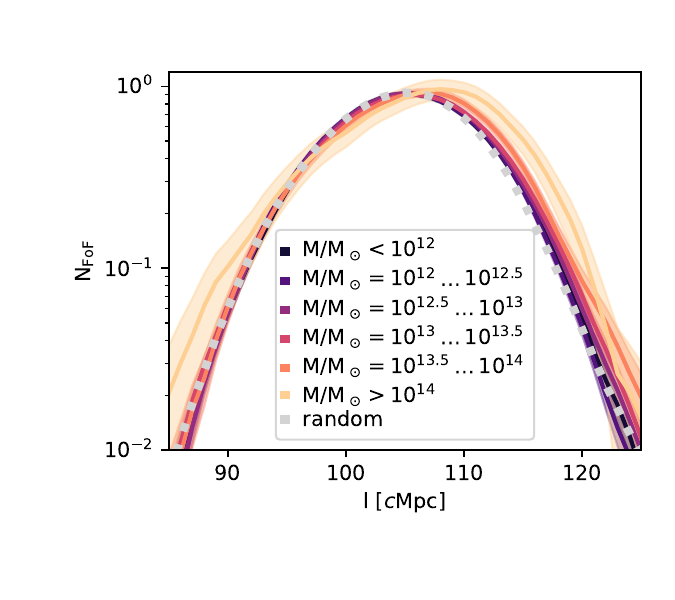}
\caption{Average number of FoF groups per slice more extended than $L=700$~$c$Mpc and more anisotropic than $b/a=0.5$, as a function of linking length, l. Different colours indicate samples of different subhalo mass ranges in the {\sc FLAMINGO-10K} simulation at $z=0.7$, as indicated in the legend; the dotted grey line shows the result for random point samples. The number of groups found is a strong function of the linking length and the value of 95~$c$Mpc that maximises the appearance of the ``Giant Arc'' is close to the value that maximises  the number of groups found both in the simulation and in random point patterns. The number of groups in {\sc FLAMINGO-10K} is very similar to that in random point patterns for all mass ranges, except for a slight excess when considering only the most massive haloes. \label{fig:linking}}
\end{figure}

\subsection{Structure persistence}
The greater number density of possible \Mgii{} absorbers in {\sc FLAMINGO-10K} compared to the \cite{Lopez-2022} sample allows us to construct independent samples of the full population. This enables us to both extend our statistics and test whether any structure identified in one subsample persists in other subsamples. Structures identified from one sample that persist in other samples of the same volume indicate a real underlying structure, while structures that are only identified in one subsample indicate a spurious pattern.

In Figure~\ref{fig:subsamples}, we show the probability density function (PDF, top) and cumulative distribution function (CDF, bottom) of the length of the longest structures satisfying $b/a < 0.5$ found within random subsamples in 21 non-overlapping slices of {\sc FLAMINGO-10K}. We find that the distributions for the length of the longest structures per slice are well approximated by a single log-normal distribution.

Almost every sample of every slice contains at least one structure longer than $\approx 400$~$c$Mpc, while the typical (median) length of the longest structure found in one random sample of any slice in {\sc FLAMINGO-10K} is $\approx 600$~$c$Mpc. Approximately $19\%$ and $11\%$ of samples contain at least one structure longer than the lower and upper limits for the extent of the ``Giant Arc'', respectively. Even accounting for the $\approx 3 \times$ larger volume compared to that examined by \cite{Lopez-2022}, thin structures as extended as the ``Giant Arc'' are very common in {\sc FLAMINGO-10K}.

We also find that, in any given slice, different random subsamples of the underlying subhalo population lead to completely different structures, but the PDFs of maximum lengths across different slices are very similar. ``Giant-Arc'' analogues are not only common, but largely spurious, rather than real underlying structures.

\subsection{Sensitivity to the linking length}
The number of structures identified with a Friends-of-Friends algorithm is expected to depend on the linking length. In Figure~\ref{fig:linking}, we show the average number of ``Giant Arc'' analogues per slice, either in samples of subhaloes of different masses in {\sc FLAMINGO-10K} at $z=0.7$  (coloured lines), or in random point samples (grey), as a function of linking length. We find that the number of structures is very sensitive to the linking length, and that for all but the highest subhalo masses, the number found in {\sc FLAMINGO-10K} is very similar to that in random point samples. Notably, the linking length originally chosen to detect the ``Giant Arc'' by \cite{Lopez-2022}, $\mathrm{l} = 95$~$c$Mpc, is close to the one that maximises the number of structures identified in both the simulation and in random point patterns.

\subsection{Overdensities}
As shown in Figure~\ref{fig:gallery}, structures resembling the ``Giant Arc'' in {\sc FLAMINGO-10K} often have point overdensities, $\delta_p$, of order unity, and comparable to the stated overdensity of the ``Giant Arc'' ($\delta_p = 0.9$). Physical overdensities of order unity on these scales would certainly be inconsistent with the $\Lambda$CDM model. However, for the same structures, we find matter overdensities, $\delta_m$, of at most a few percent, and in some cases, even slight underdensities.

\begin{figure}
\centering
\includegraphics[width=8.5cm, trim={1cm 1.2cm 0 1.cm},clip]{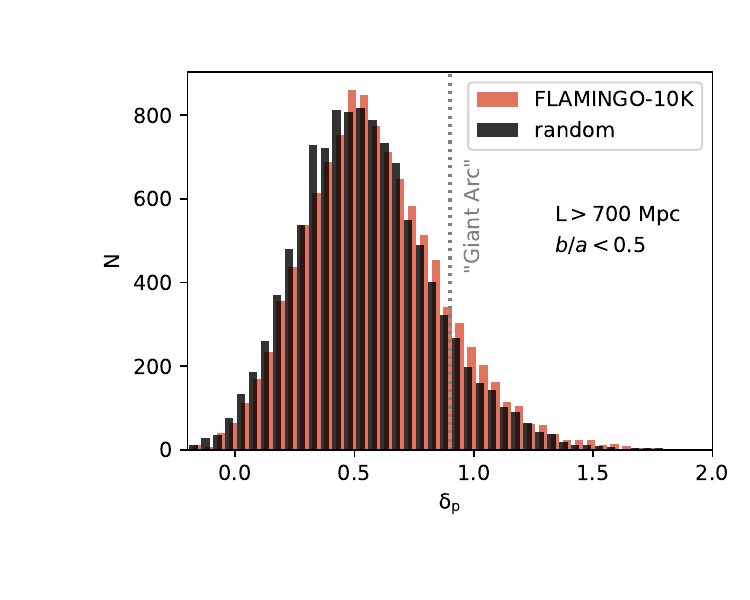}
\includegraphics[width=8.5cm, trim={1cm 1.2cm 0 1.cm},clip]{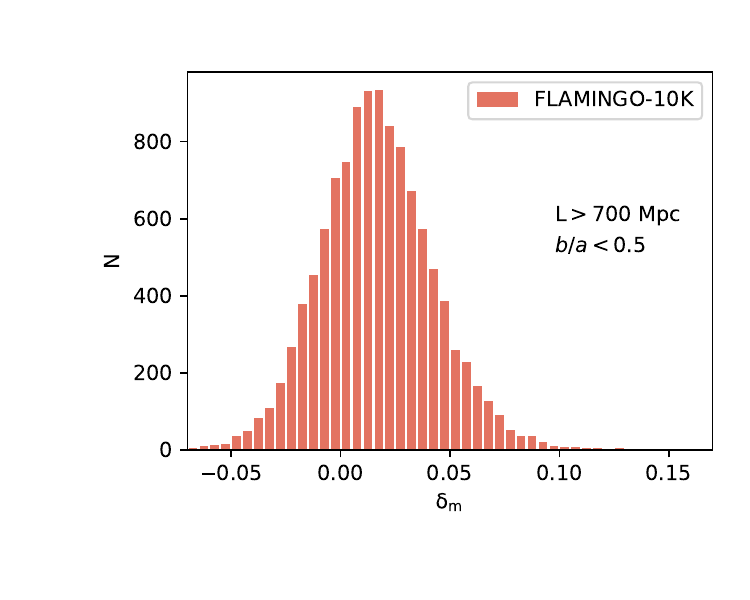}
\caption{Distribution of overdensities associated with giant structures ($L > 700$~$c$Mpc). The top panel shows the point overdensities of groups in samples of {\sc FLAMINGO-10K} at $z=0.7$ and in random point patterns. Structures in {\sc FLAMINGO-10K} are slightly more overdense than those in random patterns, but the reported point overdensity of the ``Giant Arc'', $\delta_p = 0.9$ is entirely consistent with both {\sc FLAMINGO-10K} ($87^\mathrm{th}$ percentile) and random patterns ($89^\mathrm{th}$ percentile). The bottom panel shows the matter overdensities, $\delta_m$, of the same groups in {\sc FLAMINGO-10K}. The matter overdensities are much lower, with a median of $\delta_m = 0.02$.
\label{fig:overdensity}}
\end{figure}

In the top panel of Figure~\ref{fig:overdensity}, we show histograms of the point overdensities, $\delta_p$, of extended structures in {\sc FLAMINGO-10K} and in random samples. Structures in {\sc FLAMINGO-10} are very slightly denser than those in random point patterns. Both sets are clearly overdense, and the point overdensity of the ``Giant Arc'' is not exceptional, ranking in the $87^\mathrm{th}$ and $89^\mathrm{th}$ percentiles among structures in {\sc FLAMINGO-10K} and random patterns, respectively. This contradicts the statement by \cite{Lopez-2022}, who found that its convex hull overdensity represents a ``4.5 $\sigma$ outlier''. The difference most likely arises from the fact that, while \citeauthor{Lopez-2022} considered the density of convex hulls enclosing $N$ points randomly distributed within the slice, we consider the volumes that enclose $N$ points already identified as belonging to Friends-of-Friends structures of extent similar to the ``Giant Arc''. Groups selected like the  ``Giant Arc'', with a linking length below the mean interparticle distance, are overdense almost by definition.

In the bottom panel of Figure~\ref{fig:overdensity}, we show the distribution of {\it matter} overdensities, $\delta_m$, for the same structures in {\sc FLAMINGO-10K}. Despite point overdensities of order unity, the average matter overdensity in these structures is only slightly above zero. Our results strongly suggest that this also applies to the region of the Universe that contains the ``Giant Arc''.

\subsection{Scale-and time dependent number of structures}
\begin{figure}
\centering
\includegraphics[width=8.7cm, trim={1.7cm 1.1cm 0.cm 0.cm},clip]{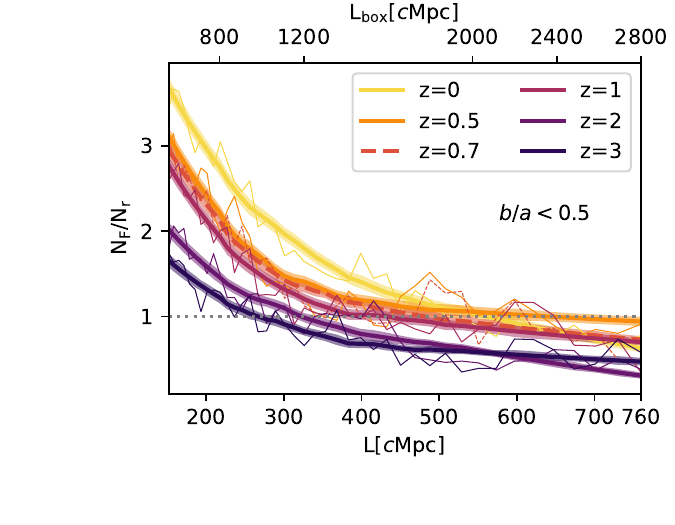}
\caption{Number of groups more extended than $8 \times$ the linking length and more anisotropic than $b/a = 0.5$ in slices of the {\sc FLAMINGO-10K} simulation at different redshifts, normalised by the corresponding number in random point patterns. Thin lines show raw measurements, thick lines are interpolated. As the length of the slice ($\mathrm{L_{box}}$, top x-axis) is varied, the linking length, l, and minimum structure size, ($\mathrm{L}$, bottom x-axis) , are scaled proportionally. As expected in the \LCDM\ paradigm, structures grow hierarchically over time. At $z=3$, {\sc FLAMINGO-10K} shows an excess of structures up to $\mathrm{L} \approx 300$~$c$Mpc, while at $z=0$, the excess extends up to $\mathrm{L}\approx 500$~$c$Mpc.
\label{fig:growth}}
\end{figure}

In Figure~\ref{fig:growth}, we show the number of extended structures of different scales and at different redshifts in slices of the {\sc FLAMINGO-10K} simulation, normalised by the corresponding numbers in random point patterns. For consistency, for the largest slices, we choose the default parameters for the volume, $2800 \times 2800 \times 338$~$c$Mpc$^3$, number of points, $N=1588$, and linking length, $l=95$~$c$Mpc, but note that these provide a rather arbitrary anchor point.

To identify structures at different scales, we set a minimum extent of $8 \times l$, corresponding to $760$~$c$Mpc for the largest slice ($\approx 20\%$ of slices contain such structures). For all smaller scales, we change all three dimensions of the slice and the linking length proportionally, maintaining a constant geometry, constant number of points, and constant ratio of the linking length to the mean interparticle separation. In this setup, the average number of structures found in random point patterns is constant and serves as a normalisation.

In the paradigm of hierarchical structure formation, the number of structures is both time- and scale dependent. As expected, we find relatively more structures at later times (lower $z$) and on smaller scales in {\sc FLAMINGO-10K}. However, even where the number of structures found in the simulation is the same or below that of random samples, it never decreases to zero: the \LCDM model predicts structures at all times and on all scales.

\section{Summary} \label{sec:summary}
Observed patterns like the ``Giant Arc'' do not contradict the $\Lambda$CDM paradigm. Applying the same detection algorithm with the same parameters assumed by \cite{Lopez-2022} for their sample of \Mgii{} absorbers to corresponding samples of subhaloes at the same redshift, the {\sc FLAMINGO-10K} simulation contains many structures that are as extended, thin, and appear as overdense as the ``Giant Arc''. Its reported overdensity holds no additional significance: {\sc FoF}-groups resembling the ``Giant Arc'' naturally possess a large points overdensity, but this does not translate into an underlying matter overdensity. Considering that the sample of \Mgii{} absorbers only represents one of many possible, extremely sparse samples of the underlying galaxy population, and that similar structures are found in random point samples, there is no reason to believe that the ``Giant Arc'' traces any underlying structure in the Universe.

We hope that our results will dispel the misconception that no inhomogeneity can be found in the standard model Universe beyond some finite size. Instead, any given realisation of the isotropic universe comprises a time- and scale-dependent population of structures from which patterns can be identified on any scale.

Structures in {\sc FLAMINGO-10K} that resemble the ``Giant Arc'' are not collapsed, gravitationally bound, or even particularly overdense. Our results strongly suggest that the same applies to the ``Giant Arc'' and other similarly large agglomerations. Without these physical characteristics commonly associated with cosmic structures, it may be more appropriate to term them ``patterns'' rather than structures.

Quantifying the frequency of direct counterparts to the  ``Giant Arc'' has been complicated by the fact that both the precise volume and the parameters of the identification algorithm were defined post-hoc, with the explicit goal of obtaining the clearest detection of a previously identified structure. For consistency, we adopt the same parameters, but this strategy clearly entails a significant look-elsewhere effect. However, the fact that we find so many counterparts in {\sc FLAMINGO-10K} even with these particular parameters clearly demonstrates that its discovery is fully consistent with the standard model.

\section*{Data Availability Statement}
The script used to produce all figures and numbers presented in this work is available at https://github.com/TillSawala/GiantArc. Access to the underlying simulation data will be provided on reasonable request to the authors.

\section*{Acknowledgements}
TS and GR acknowledge support from Academy of Finland grant 354905, and TS and MT acknowledge support from Academy of Finland grant 339127. TS and CSF acknowledge support from European Research Council (ERC) Advanced Grant DMIDAS (GA 786910). GR acknowledges support from European Research Council (ERC) Consolidator Grant KETJU (GA 818930). This work used the DiRAC@Durham facility managed by the ICC, with support from BEIS via STFC capital grants ST/K00042X/1, ST/P002293/1, ST/R002371/1 and ST/S002502/1, and STFC operations grant ST/R000832/1. This work used the DiRAC Memory Intensive service (Cosma8) at the University of Durham, which is part of the STFC DiRAC HPC Facility (www.dirac.ac.uk). Access to DiRAC resources was granted through a Director’s Discretionary Time allocation in 2023/24, under the auspices of the UKRI-funded DiRAC Federation Project. We used open source software, including Matplotlib \citep{matplotlib-paper}, SciPy \citep{SciPy}, Scikit-learn \citep{scikit-learn} and NumPy \citep{numpy-paper}.



\bibliographystyle{mnras} \bibliography{paper}


\bsp	
\label{lastpage}
\end{document}